\documentclass [twocolumn,10.5pt]{article}
\title{Comment on "Observation of the e/3 fractionally charged Laughlin quasiparticles" and "Direct Observation of Fractional Charge".}

\begin{document}
\maketitle
In two recent manuscripts [1,2] reporting measurements of shot noise in the fractional quantum Hall regime, inaccurate statements were made concerning our previously published [3] observation of fractional charge in antidot tunneling experiments. Particularly, it is claimed in [2] that "A more recent experiment by Goldman and Su [3] was reproduced and interpreted differently by Franklin et al. [4]". Ref. 1 states: "In a recent beautiful experiment using an antidot at nu = 1/3 [3], the period of the polarisation charge on the control back-gate was found accurately e/3 = e*. In a similar report, it has been argued that such equilibrium conductance measurements only accurately probe the fractional filling of the ground state [4]".

However, Franklin et al. did not reproduce our experiment as far as the measurement of charge is concerned, neither do they claim this in [4] - for the simple reason that their samples did not have a macroscopic global remote gate (our global backgate), which produces a uniform electric field acting on the 2D electrons, which is absolutely essential for our charge measurements. Franklin et al. had a small front gate which defined the antidot itself by creating spatially strongly nonuniform electric field.

Franklin et al. did have a vague discussion regarding our experiments; it is probably naive to regard their comments at that time as definitive and final conclusions. In particular, it should be clear that our experiment cannot measure "filling of the ground state", since it does measure charge in Coulombs (not in units of the electron charge), while filling factor is a unitless number. In our experiment the direct measurement is that of the coupling of a particle to uniform electric field, that is of charge of that particle. The experiment does measure charge of one particle per (many-body) state; indeed, using this charge, in units of the assumed to be known electron charge, and assuming the ratio of flux quanta per state, one may obtain the unitless filling factor. Thus, to deduce the filling factor of the condensate (which is not a directly observable quantity), one has to devide the measured charge by the charge of one electron, assumed to be known. 

We also note that our interpretation of direct charge measurement is generally accepted in the theoretical community as essentially correct. Thus, the first direct observation and accurate measurement of fractional charge e/3 and e/5 has been reported by us in [3,5].
\\
\\
Vladimir J. Goldman\\
Department of Physics\\
SUNY at Stony Brook\\
Stony Brook, NY 11794-3400\\
\\
1. L. Saminadayar et al., cond-mat/97063007\\
2. R. de-Picciotto et al., cond-mat/9707289\\
3. V.J. Goldman and B.Su, Science 267, 1010 (1995).\\ 
4. J.D.F. Franklin et al., Surf. Sci. 361, 17 (1996).\\
5. V.J. Goldman, Surf. Sci. 361, 1 (1996).\\
\end{document}